\documentclass[runningheads]{llncs}
\usepackage{graphicx}
\usepackage{multirow}
\usepackage{booktabs}
\usepackage{bbding}
\usepackage{hyperref, color}

\usepackage{amsmath}
\usepackage{algorithm}
\usepackage{algorithmic}
\usepackage[switch]{lineno}
\usepackage{threeparttable}
\usepackage{rotating}

\begin{document}
\title{SAM Meets Robotic Surgery: An Empirical Study on Generalization, Robustness and Adaptation}
\titlerunning{SAM Meets Robotic Surgery}
%
\author{An Wang\inst{1~\star} 
\and Mobarakol Islam\inst{2} 
\thanks{An Wang and Mobarakol Islam are co-first authors.} 
\and Mengya Xu\inst{3}
\and Yang Zhang\inst{4}
\and Hongliang Ren\inst{1,3} 
\thanks{Corresponding author.}}

%
\authorrunning{A. Wang et al.}
%
\institute{Dept. of Electronic Engineering, Shun Hing Institute of Advanced Engineering (SHIAE), The Chinese University of Hong Kong, Hong Kong SAR, China
\and Dept. of Medical Physics and Biomedical Engineering, Wellcome/EPSRC Centre for Interventional and Surgical Sciences (WEISS), University College London, London, UK
\and Dept. of Biomedical Engineering, National University of Singapore, Singapore
\and School of Mechanical Engineering, Hubei University of Technology, Wuhan, China\\
\email{wa09@link.cuhk.edu.hk, mobarakol.islam@ucl.ac.uk, mengya@u.nus.edu, yzhangcst@hbut.edu.cn, hlren@ee.cuhk.edu.hk}
}
\maketitle              
\begin{abstract}
The Segment Anything Model (SAM) serves as a fundamental model for semantic segmentation and demonstrates remarkable generalization capabilities across a wide range of downstream scenarios. In this empirical study, we examine SAM's robustness and zero-shot generalizability in the field of robotic surgery. We comprehensively explore different scenarios, including prompted and unprompted situations, bounding box and points-based prompt approaches, as well as the ability to generalize under corruptions and perturbations at five severity levels. Additionally, we compare the performance of SAM with state-of-the-art supervised models. We conduct all the experiments with two well-known robotic instrument segmentation datasets from MICCAI EndoVis 2017 and 2018 challenges. Our extensive evaluation results reveal that although SAM shows remarkable zero-shot generalization ability with bounding box prompts, it struggles to segment the whole instrument with point-based prompts and unprompted settings. Furthermore, our qualitative figures demonstrate that the model either failed to predict certain parts of the instrument mask (e.g., jaws, wrist) or predicted parts of the instrument as wrong classes in the scenario of overlapping instruments within the same bounding box or with the point-based prompt. In fact, SAM struggles to identify instruments in complex surgical scenarios characterized by the presence of blood, reflection, blur, and shade. Additionally, SAM is insufficiently robust to maintain high performance when subjected to various forms of data corruption. We also attempt to fine-tune SAM using Low-rank Adaptation (LoRA) and propose SurgicalSAM, which shows the capability in class-wise mask prediction without prompt. Therefore, we can argue that, without further domain-specific fine-tuning, SAM is not ready for downstream surgical tasks.
\end{abstract}
\section{Introduction}
\label{sec:introduction}


Segmenting surgical instruments and tissue poses a significant challenge in robotic surgery, as it plays a vital role in instrument tracking and position estimation within surgical scenes. Nonetheless, current deep learning models often have limited generalization capacity as they are tailored to specific surgical sites. Consequently, it is crucial to develop generalist models that can effectively adapt to various surgical scenes and segmentation objectives to advance the field of robotic surgery~\cite{seenivasan2023surgicalgpt}. Recently, segmentation foundation models have made great progress in the field of natural image segmentation. The segment anything model (SAM)~\cite{kirillov2023segment}, which has been trained on more than one billion masks, exhibits remarkable proficiency in generating precise object masks using various prompts such as bounding boxes and points. SAM stands as the pioneering and most renowned foundation model for segmentation. Whereas, several works have revealed that SAM can fail on common medical image segmentation tasks~\cite{deng2023segment,hu2023sam,he2023accuracy,ma2023segment}. This is not surprising or unexpected since SAM's training dataset primarily comprises natural image datasets. 
Consequently, it raises the question of enhancing SAM's strong feature extraction capability for medical image tasks. Med SAM Adapter~\cite{wu2023medical} utilizes medical-specific domain knowledge to improve the segmentation model through a simple yet effective adaptation technique. SAMed~\cite{zhang2023customized} has applied a low-rank-based finetuning strategy to the SAM image encoder, as well as prompt encoder and mask decoder on the medical image segmentation dataset. 

However, evaluating the performance of SAM in the context of surgical scenes remains an insufficiently explored area that has the potential for further investigation. This study uses two publicly available robotic surgery datasets to assess SAM's generalizability under different settings, such as bounding box and point-prompted. Moreover, we have examined the possibility of fine-tuning SAM through Low-rank Adaptation (LoRA) to examine its capability to predict masks for different classes without prompts. Additionally, we have analyzed SAM's robustness by assessing its performance on synthetic surgery datasets, which contain various levels of corruption and perturbations.

\section{Experimental Settings}
\noindent\textbf{Datasets.}
We have employed two classical datasets in endoscopic surgical instrument segmentation, i.e., EndoVis17~\cite{allan20192017} and EndoVis18~\cite{allan20202018}. 
For the EndoVis17 dataset, unlike previous works~\cite{shvets2018automatic,gonzalez2020isinet,jin2019incorporating} which conduct 4-fold cross-validation for training and testing on the 8$\times$225-frame released training data, we report SAM's performance directly on all eight sequences (1-8). For the EndoVis18 dataset, we follow the dataset split in ISINet~\cite{gonzalez2020isinet}, where sequences 2, 5, 9, and 15 are utilized for evaluation. 

\noindent\textbf{Prompts.} 
The original EndoVis datasets~\cite{allan20192017,allan20202018} do not have bounding boxes or point annotations. 
We have labeled the datasets with bounding boxes for each instrument, associated with corresponding class information. Additionally, regarding the single-point prompt, we obtain the center of each instrument mask by simply computing the moments of the mask contour. Since SAM~\cite{kirillov2023segment} only predicts binary segmentation masks, for instrument-wise segmentation, the output instrument labels are assigned inherited from the input prompts. 

\noindent\textbf{Metrics.} 
The IoU and Dice metrics from the EndoVis17~\cite{allan20192017} challenge\footnote{\url{https://github.com/ternaus/robot-surgery-segmentation}} is used. Specifically, only the classes presented in a frame are considered in the calculation for instrument segmentation.

\noindent\textbf{Comparison methods.}
We have involved several classical and recent methods, including the vanilla UNet~\cite{ronneberger2015u}, TernausNet~\cite{shvets2018automatic}, MF-TAPNet~\cite{jin2019incorporating}, Islam et al.~\cite{islam2019real}, Wang et al.~\cite{wang2022rethinking}, ST-MTL~\cite{islam2021st}, S-MTL~\cite{seenivasan2022global}, AP-MTL~\cite{islam2020ap}, ISINet~\cite{gonzalez2020isinet}, TraSeTR~\cite{zhao2022trasetr}, and S3Net~\cite{baby2023forks} for surgical binary and instrument-wise segmentation. The ViT-H-based SAM~\cite{kirillov2023segment} is employed in all our investigations except for the finetuning experiments. Note that we cannot provide an absolutely fair comparison because existing methods do not need prompts during inference.

\begin{table}[!t]
  \centering
  \caption{Quantitative comparison of binary and instrument segmentation on EndoVis17 and EndoVis18 datasets. The best and runner-up results are shown in bold and underlined.}
  \begin{threeparttable}
    \resizebox{0.95\textwidth}{!}{
    \begin{tabular}{cccccccc}
    \toprule
    \multirow{2}[2]{*}{Type} & \multirow{2}[2]{*}{Method} & \multirow{2}[2]{*}{Pub/Year(20-)} & \multirow{2}[2]{*}{Arch.} & \multicolumn{2}{c}{EndoVis17} & \multicolumn{2}{c}{EndoVis18} \\
\cmidrule{5-8}          &       &       &       & Binary IoU & Instrument IoU & Binary IoU & Instrument IoU \\
    \midrule
    \multirow{5}[2]{*}{Single-Task} & Vanilla UNet & MICCAI15 & UNet  & 75.44 & 15.80 & \underline{68.89} & - \\
          & TernausNet & ICMLA18 & UNet  & 83.60 & 35.27 & -     & 46.22 \\
          & MF-TAPNet & MICCAI19 & UNet  & 87.56 & 37.35 & -     & 67.87 \\
          & Islam et al. & RA-L19 & -     & 84.50 & -     & -     & - \\
          & ISINet & MICCAI21 & Res50 & -     & 55.62 & -     & 73.03 \\
          & Wang et al. & MICCAI22 & UNet  & -     & -     & 58.12 & - \\
    \midrule
    \multirow{5}[0]{*}{Multi-Task} & ST-MTL & MedIA21 & -     & 83.49 & -     & -     & - \\
          & AP-MTL & ICRA20 & -     & \underline{88.75} & -     & -     & - \\
          & S-MTL & RA-L22 & -     & -     & -     & -     & 43.54 \\
          & TraSeTR & ICRA22 & Res50 + Trfm & -     & 60.40 & -     & \underline{76.20} \\
          & S3Net & WACV23 & Res50 & -     & \underline{72.54} & -     & 75.81 \\
    \midrule
    \multirow{2}[0]{*}{Prompt-based} & SAM 1 Point & arxiv23 & ViT\_h   & 53.88 & 55.96\tnote{*} & 57.12 & 54.30\tnote{*} \\
          & SAM Box & arxiv23 & ViT\_h   & \textbf{89.19} & \textbf{88.20}\tnote{*} & \textbf{89.35} & \textbf{81.09}\tnote{*} \\
    \bottomrule
    \end{tabular}%
    }\end{threeparttable}
    \begin{tablenotes}
    \footnotesize
    \item{*} Categorical information directly inherits from associated prompts.
    \end{tablenotes}
  \label{tab:overall_res}%
\end{table}%

\section{Surgical Instruments Segmentation with Prompts}
\subsubsection{Implementation}
With bounding boxes and single points as prompts, we input the images to SAM~\cite{kirillov2023segment} to get the predicted binary masks for the target objects. Because SAM~\cite{kirillov2023segment} can not provide consistent categorical information. We compromise to use the class information from the bounding boxes directly. In this way, we derive instrument-wise segmentation while bypassing the possible errors from misclassifications, an essential factor affecting instrument-wise segmentation accuracy.  

\subsubsection{Results and Analysis}
\begin{figure*}[!t]
  \centering
  \includegraphics[width=0.65\linewidth]{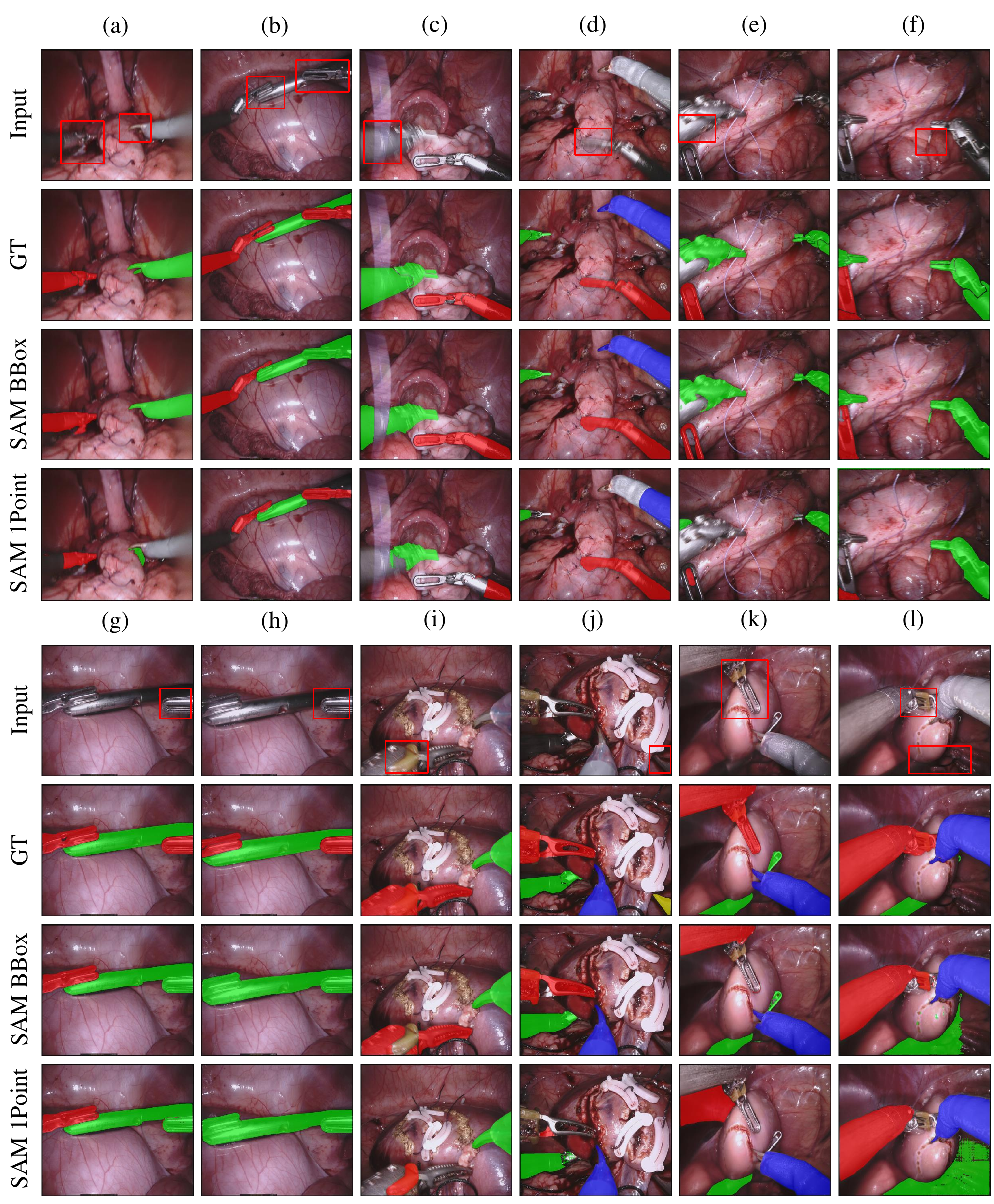}
  \caption{Qualitative results of SAM on various challenging frames. Red rectangles highlight the typical challenging regions which cause unsatisfactory predictions.}
  \label{fig:quali}
\end{figure*}

As shown in Table~\ref{tab:overall_res}, with bounding boxes as prompts, SAM~\cite{kirillov2023segment} outperforms previous unprompted supervised methods in binary and instrument-wise segmentation on both datasets. 
However, with single points as prompts, SAM~\cite{kirillov2023segment} degrades a lot in performance, indicating its limited ability to segment surgical instruments from weak prompts. This reveals the performance of the SAM closely relies on prompt quality.
For complicated surgical scenes, SAM~\cite{kirillov2023segment} still struggles to produce accurate segmentation results, as shown in columns (a) to (l) of Fig.~\ref{fig:quali}. Typical challenges, including shadows (a), motion blur (d), occlusion (b, g, h), light reflection (c), insufficient light (j, l), over brightness (e), ambiguous suturing thread (f), instrument wrist (i), and irregular instrument pose (k), all lead to unsatisfied segmentation performance.

\begin{table}[!t]
  \centering
  \caption{Quantitative results on various corrupted EndoVis18 validation data.}
    \resizebox{\textwidth}{!}{
    \begin{tabular}{cc|cccc|ccccc|ccccc|cccc}
    \toprule
    \multirow{2}[0]{*}{Task} & \multirow{2}[0]{*}{Severity} & \multicolumn{4}{c|}{Noise}    & \multicolumn{5}{c|}{Blur}             & \multicolumn{5}{c|}{Weather}          & \multicolumn{4}{c}{Digital} \\
          &       & Gaussian & Shot  & Impulse & Speckle & Defocus & Glass & Motion & Zoom  & Gaussian & Snow  & Frost & Fog   & Bright & Spatter & Contrast & Pixel & JPEG  & Saturate \\
    \midrule
    \multirow{6}[2]{*}{\begin{sideways}Binary\end{sideways}} & \multicolumn{1}{c}{0} & \multicolumn{18}{c}{89.35} \\
\cmidrule{2-20}          & 1     & 77.69 & 80.18 & 80.43 & 83.28 & 82.01 & 80.53 & 82.99 & 80.30 & 85.40 & 84.08 & 83.12 & 85.38 & 87.43 & 86.69 & 85.76 & 81.12 & 58.77 & 86.64 \\
          & 2     & 73.92 & 76.07 & 76.15 & 81.65 & 80.21 & 79.20 & 80.22 & 77.55 & 81.69 & 80.69 & 80.34 & 84.65 & 87.27 & 84.21 & 84.90 & 79.32 & 56.04 & 84.85 \\
          & 3     & 69.21 & 71.74 & 73.02 & 77.74 & 76.96 & 72.64 & 75.50 & 75.27 & 78.31 & 79.58 & 78.90 & 83.62 & 87.23 & 82.50 & 83.36 & 73.81 & 56.25 & 86.84 \\
          & 4     & 63.80 & 65.41 & 67.29 & 75.28 & 73.79 & 72.38 & 69.60 & 73.22 & 75.23 & 76.33 & 78.38 & 82.28 & 87.06 & 83.12 & 77.12 & 70.82 & 57.59 & 83.21 \\
          & 5     & 57.07 & 60.61 & 61.61 & 71.83 & 69.85 & 69.59 & 66.25 & 71.58 & 66.96 & 77.66 & 76.82 & 78.84 & 86.43 & 79.62 & 66.58 & 68.55 & 56.77 & 81.26 \\
    \midrule
    \multirow{6}[2]{*}{\begin{sideways}Instrument\end{sideways}} & \multicolumn{1}{c}{0} & \multicolumn{18}{c}{81.09} \\
\cmidrule{2-20}          & 1     & 69.51 & 71.83 & 72.25 & 74.82 & 73.64 & 72.13 & 74.33 & 71.41 & 76.79 & 75.40 & 74.42 & 76.82 & 79.16 & 78.24 & 77.17 & 72.94 & 54.86 & 78.27 \\
          & 2     & 66.06 & 68.09 & 68.53 & 73.19 & 71.74 & 71.02 & 71.46 & 68.85 & 73.15 & 72.13 & 71.65 & 76.14 & 79.00 & 75.54 & 76.22 & 71.55 & 52.23 & 76.61 \\
          & 3     & 62.01 & 64.44 & 65.89 & 69.75 & 68.74 & 64.97 & 67.13 & 67.12 & 70.08 & 70.97 & 70.21 & 75.01 & 78.90 & 73.70 & 74.67 & 66.83 & 51.63 & 78.39 \\
          & 4     & 57.28 & 59.12 & 61.03 & 67.82 & 65.87 & 64.87 & 62.15 & 65.18 & 67.23 & 68.43 & 69.79 & 73.73 & 78.73 & 74.24 & 69.48 & 63.99 & 51.88 & 74.91 \\
          & 5     & 51.56 & 55.16 & 55.86 & 64.76 & 62.43 & 62.23 & 59.26 & 63.96 & 60.60 & 69.33 & 68.32 & 70.45 & 78.19 & 70.72 & 61.14 & 61.79 & 51.01 & 73.35 \\
    \bottomrule
    \end{tabular}%
    }
  \label{tab:corruption}%
\end{table}%

\section{Robustness under Data Corruption}
\subsubsection{Implementation}
Referring to the robustness evaluation benchmark~\cite{hendrycks2018benchmarking}, we have evaluated SAM~\cite{kirillov2023segment} under 18 types of data corruptions at 5 severity levels following the official implementations\footnote{\url{https://github.com/hendrycks/robustness}} with box prompts. Note that the \textit{Elastic Transformation} has been omitted to avoid inconsistency between the input image and associated masks. The adopted data corruption can be allocated into four distinct categories of \textit{Noise}, \textit{Blue}, \textit{Weather}, and \textit{Digital}.

\begin{figure*}[!t]
  \centering
  \includegraphics[width=0.65\linewidth]{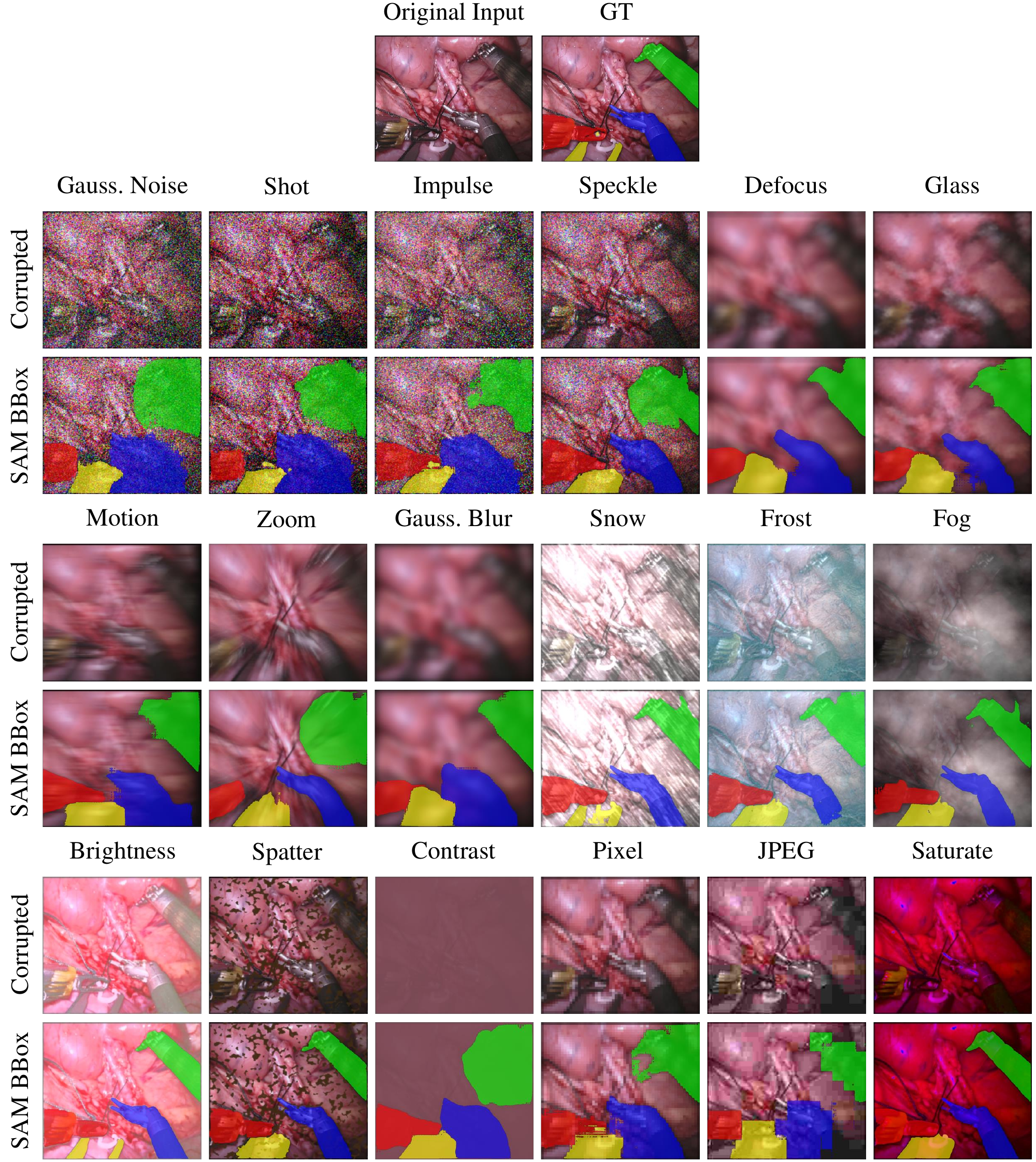}
  \caption{Qualitative results of SAM under 18 data corruptions of level-5 severity.}
  \label{fig:corrupt}
\end{figure*}

\subsubsection{Results and Analysis}
The severity of data corruption is directly proportional to the degree of performance degradation in SAM~\cite{kirillov2023segment}, as depicted in Table~\ref{tab:corruption}.
The robustness of SAM~\cite{kirillov2023segment} may be influenced differently depending on the nature of the corruption present. However, in most scenarios, SAM's performance diminishes significantly. Notably, \textit{JPEG Compression} and \textit{Gaussian Noise} have the greatest impact on segmentation performance, whereas \textit{Brightness} has a negligible effect. 
Figure~\ref{fig:corrupt} presents one exemplar frame in its original state alongside various corrupted versions at a severity level of 5. We can observe that SAM~\cite{kirillov2023segment} suffers significant performance degradation in most cases.

\section{Automatic Surgical Scene Segmentation}
\subsubsection{Implementation}
Without prompts, SAM~\cite{kirillov2023segment} can also facilitate automatic mask generation (AMG) for the entire image. 
For naive investigation of the automatic surgical scene segmentation results, we use the default parameters from the official implementation\footnote{\url{https://github.com/facebookresearch/segment-anything}} without further tuning. The colors of each segmented mask are randomly assigned because SAM~\cite{kirillov2023segment} only generates binary masks for each object. 

\subsubsection{Results and Analysis}
\begin{figure*}[!t]
  \centering
  \includegraphics[width=0.8\linewidth]{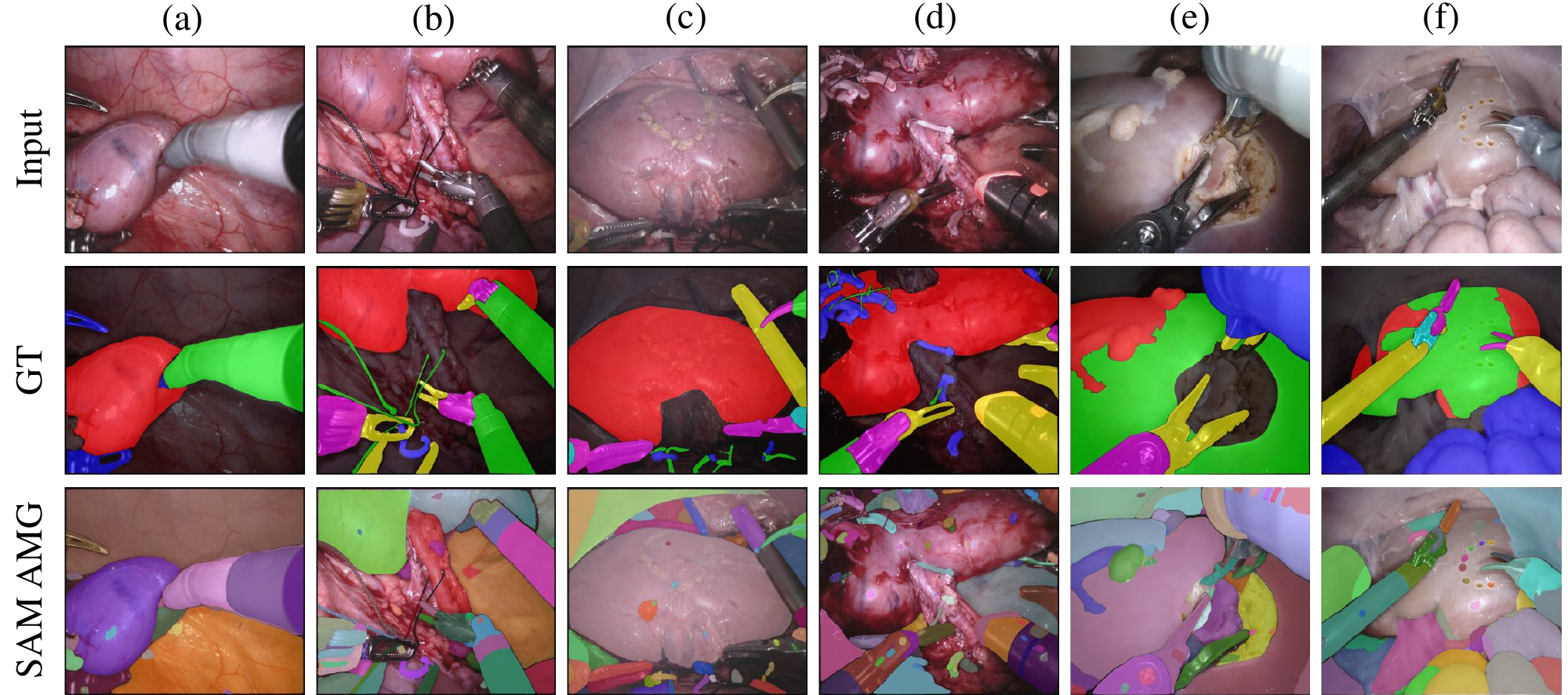}
  \caption{Unprompted automatic mask generation for surgical scene segmentation.}
  \label{fig:amg}
\end{figure*}

As shown in Fig.~\ref{fig:amg}, in surgical scene segmentation of EndoVis18~\cite{allan20202018} data, SAM~\cite{kirillov2023segment} can produce promising results on simple scenes like columns (a) and (f). But it encounters difficulties when applied to more complicated scenes, as it struggles to differentiate between the entirety of instrument articulating parts accurately and to identify discrete tissue structures as interconnected units. As a foundation model, SAM~\cite{kirillov2023segment} still lacks comprehensive awareness of objects' semantics, especially in downstream domains like surgical scenes. 

\section{Parameter-efficient Finetuning with Low-rank Adaptation}
With the rapid emergence of foundational and large AI models, utilizing the pretrained models effectively and efficiently for downstream tasks has attracted increasing research interest. 
Although SAM~\cite{kirillov2023segment} has shown decent segmentation performance with prompts and can cluster objects in surgical scenes, we seek to finetune and adapt it to make it capable of traditional unprompted multi-class segmentation pipeline - take one image as input only, and predict its segmentation mask with categorical labels.

\begin{figure*}[!t]
  \centering
  \includegraphics[width=0.8\linewidth]{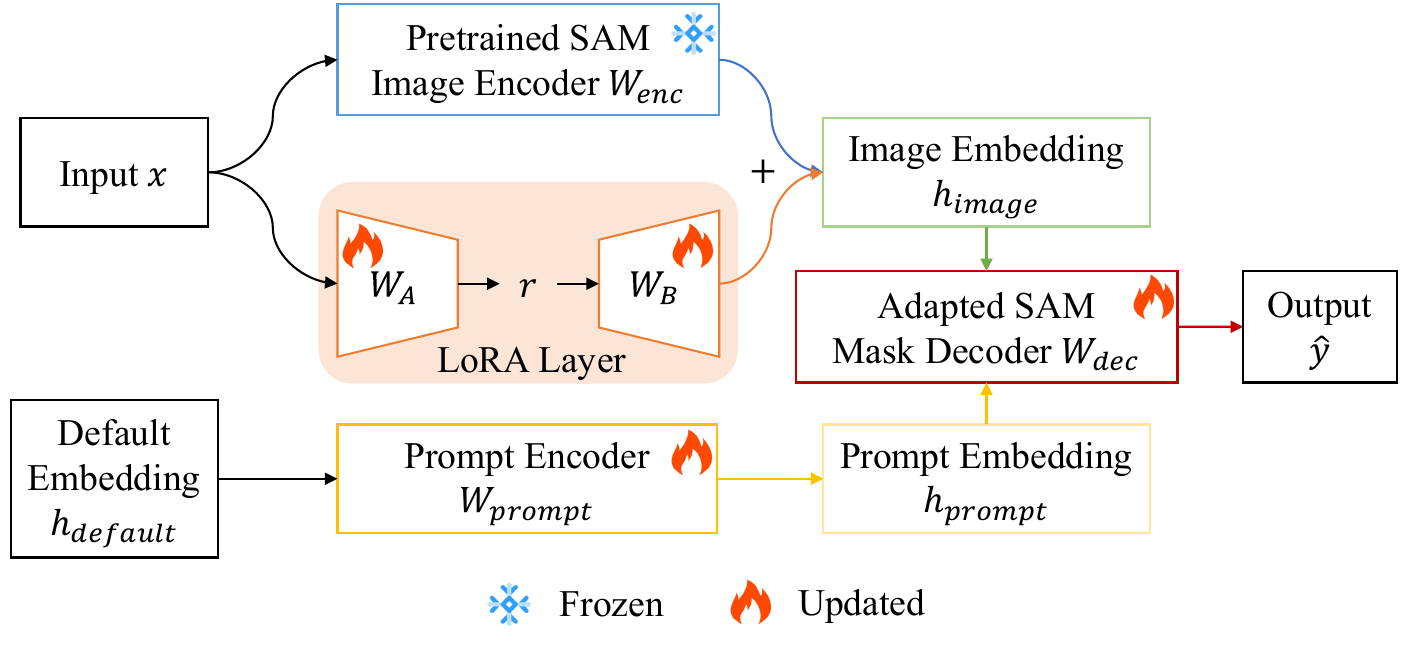}
  \caption{Overall architecture of our SurgicalSAM.}
  \label{fig:sam_lora}
\end{figure*}

\subsubsection{Implementation}
To efficiently finetune SAM~\cite{kirillov2023segment} and enable it to support multi-class segmentation without relying on prompts, we consider utilizing the strategy of Low-rank Adaptation (LoRA)~\cite{hu2021lora} and also adapting the original mask decoder to output categorical labels. 
Taking inspiration from SAMed~\cite{zhang2023customized}, we implement a modified architecture as shown in Fig.~\ref{fig:sam_lora}, whereby the pretrained SAM image encoder maintains its frozen weights $W_{enc}$ during finetuning while additional light-weight LoRA layers are incorporated for updating purposes. In this way, we can not only leverage the exceptional feature extraction ability of the original SAM encoder, but also gradually capture the surgical data representations and store the domain-specific knowledge in the LoRA layers parameter-efficiently. We denote this modified architecture as ``SurgicalSAM''.
With an input image $x$, we can derive the image embedding $h_{image}$ following
\begin{equation}
    h_{image} = W_{enc}x + \Delta W x,
\end{equation}
where $\Delta W$ is the weight update matrix of LoRA layers. Then we can decompose $\Delta W$ into two smaller matrices: $\Delta W = W_A W_B$, where $W_A$ and $W_B$ are $A \times r$ and $r \times B$ dimensional matrices, respectively. $r$ is a hyper-parameter that specifies the rank of the low-rank adaptation matrices.
To maintain a balance between model complexity, adaptability, and the potential for underfitting or overfitting, we empirically set the rank $r$ of $W_A$ and $W_B$ in the LoRA layers to 4. 

During the unprompted automatic mask generation (AMG), the original SAM uses fixed default embeddings $h_{default}$ for the prompt encoder with weights $W_{prompt}$. We adopted this strategy and updated the lightweight prompt encoder during finetuning, as shown in Fig.~\ref{fig:sam_lora}. In addition, we modified the segmentation head of the mask decoder $W_{dec}$ to allow for the production of predictions for each semantic class. In contrast to the binary ambiguity prediction of the original mask decoder of SAM, the modified decoder predicts each semantic class of $\hat{y}$ in a deterministic manner. In other words, it is capable of semantic segmentation beyond binary segmentation. 

We adopt the training split of the Endo18 dataset for finetuning and test with the validation split, as other works reported in Table~\ref{tab:overall_res}. Following SAMed~\cite{zhang2023customized}, we adopt the combination of the Cross Entropy loss $L_{CE}$ and Dice loss $L_{Dice}$ which can be expressed as 
\begin{equation}
    L = \lambda L_{Dice} + (1-\lambda) L_{CE},
\end{equation}
where $\lambda$ is a weighting coefficient balancing the effects of the two losses. We empirically set $\lambda$ as 0.8 in our experiments.
Due to resource constraints, we utilize the ViT\_b version of SAM and finetuning on two RTX3090 GPUs. 
The maximum epochs are 160, with a batch size 12 and an initial learning rate of 0.001. To stabilize the finetuning process, we apply warmup for the first 250 iterations, followed by exponential learning rate decay. 
Random flip, rotation, and crop are applied to augment the training images and avoid overfitting. The images are resized to $512 \times 512$ as model inputs. 
Besides, we use AdamW~\cite{loshchilov2017decoupled} optimizer with a weight decay of 0.1 to update model parameters.

\begin{table}[t]
  \centering
  \caption{Quantitative evaluation of SurgicalSAM under data corruption.}
    \resizebox{\textwidth}{!}{
    \begin{tabular}{c|cccc|ccccc|ccccc|cccc}
    \toprule
    \multirow{2}[2]{*}{Severity} & \multicolumn{4}{c|}{Noise}    & \multicolumn{5}{c|}{Blur}             & \multicolumn{5}{c|}{Weather}          & \multicolumn{4}{c}{Digital} \\
          & Gaussian & Shot  & Impulse & Speckle & Defocus & Glass & Motion & Zoom  & Gaussian & Snow  & Frost & Fog   & Bright & Spatter & Contrast & Pixel & JPEG  & Saturate \\
    \midrule
    \multicolumn{1}{c}{0} & \multicolumn{18}{c}{71.38} \\
    \midrule
    1     & 24.31 & 30.68 & 28.88 & 45.53 & 59.50 & 60.21 & 61.29 & 56.32 & 64.67 & 57.84 & 54.80 & 54.95 & 66.67 & 65.74 & 57.56 & 64.81 & 54.30 & 60.01 \\
    2     & 12.19 & 15.43 & 12.77 & 36.92 & 53.85 & 56.48 & 55.72 & 52.81 & 55.54 & 29.68 & 36.33 & 51.32 & 63.73 & 62.59 & 50.89 & 64.00 & 49.56 & 28.92 \\
    3     & 5.84  & 6.30  & 7.34  & 17.26 & 45.56 & 43.71 & 50.97 & 49.55 & 47.24 & 42.20 & 26.31 & 44.17 & 62.22 & 60.65 & 36.90 & 54.99 & 46.24 & 64.85 \\
    4     & 4.26  & 4.15  & 4.63  & 10.19 & 39.23 & 39.64 & 43.27 & 46.38 & 39.65 & 30.21 & 25.80 & 38.28 & 60.90 & 51.22 & 16.42 & 40.64 & 36.69 & 60.36 \\
    5     & 3.79  & 3.79  & 3.92  & 6.37  & 32.49 & 38.05 & 38.16 & 43.99 & 26.67 & 13.97 & 20.60 & 20.92 & 59.64 & 40.51 & 4.95  & 34.00 & 24.03 & 50.50 \\
    \bottomrule
    \end{tabular}%
    }
  \label{tab:lora_c}%
\end{table}%

\begin{figure}[t]
  \centering
  \includegraphics[width=0.6\linewidth]{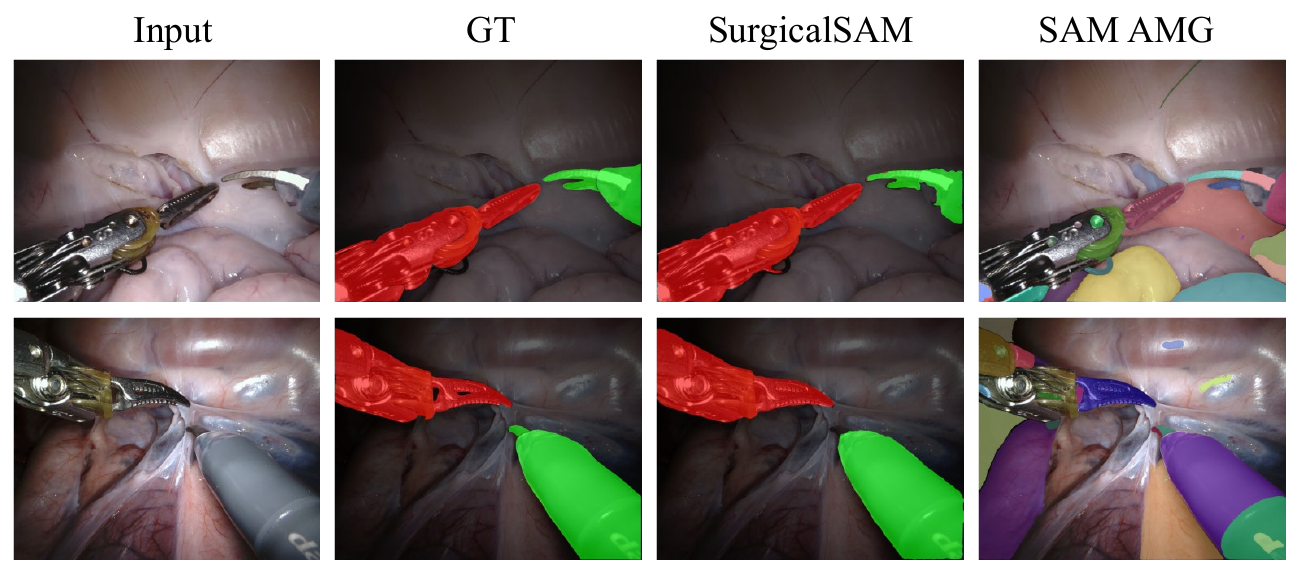}
  \caption{Qualitative comparison of our SurgicalSAM with the original SAM.} 
  \label{fig:sam_lora_quality}
\end{figure}
\subsubsection{Results and Analysis}
After naively finetuning, the SurgicalSAM model can manage the instrument-wise segmentation without reliance on prompts. With further tuning of hyper-parameters like the learning rate, the batch size, and the optimizer, SurgicalSAM can achieve \textbf{71.38\%} mIoU score on the validation split of the Endo18 dataset, which is on par with the state-of-the-art models in Table~\ref{tab:overall_res}. 
Since other methods in Table~\ref{tab:overall_res} are utilizing temporal and optical flow information as supplement~\cite{gonzalez2020isinet}, or conducting multi-task optimization~\cite{zhao2022trasetr,baby2023forks}, the results of our image-only and single-task architecture SurgicalSAM are promising. Besides, the encoder backbone we finetuned is the smallest ViT\_b due to limited computational resources. We believe the largest ViT\_h backbone can yield much better performance. Compared with the original SAM, our new architecture is of great practical significance as it can achieve semantic-level automatic segmentation. Moreover, the additionally trained parameters are only \textbf{18.28MB}, suggesting the efficiency of our finetuning strategy. 

Furthermore, we have evaluated the robustness of SurgicalSAM in the face of data corruption using the EndoVis18 validation dataset. As shown in Table~\ref{tab:lora_c}, the model's performance exhibits a significant degradation when subjected to various forms of data corruption, particularly in the case of \textit{Blur} corruption.

\section{Conclusion}\label{sec:conclusion}

In this study, we explore the robustness and zero-shot generalizability of the SAM~\cite{kirillov2023segment} in the field of robotic surgery on two robotic instrument segmentation datasets of MICCAI EndoVis 2017 and 2018 challenges, respectively. 
Extensive empirical results suggest that SAM~\cite{kirillov2023segment} is deficient in segmenting the entire instrument with point-based prompts and unprompted settings, as clearly shown in Fig.~\ref{fig:quali} and Fig.~\ref{fig:amg}. This implies that SAM~\cite{kirillov2023segment} can not capture the surgical scenes precisely despite yielding surprising zero-shot generalization ability. 
Besides, it exhibits challenges in accurately predicting certain parts of the instrument mask when there are overlapping instruments or only with a point-based prompt. 
It also fails to identify instruments in complex surgical scenarios, such as blood, reflection, blur, and shade. 
Moreover, we extensively evaluate the robustness of SAM~\cite{kirillov2023segment} with a wide range of data corruptions. As indicated by Table~\ref{tab:corruption} and Fig.~\ref{fig:corrupt}, SAM~\cite{kirillov2023segment} encounters significant performance degradation in many scenarios. 
To shed light on adapting SAM for surgical tasks, we fine-tuned the SAM using LoRA. Our fine-tuned SAM, i.e., SurgicalSAM, demonstrates the capability of class-wise mask prediction without any prompt.

As a foundational segmentation model, SAM~\cite{kirillov2023segment} shows remarkable generalization capability in robotic surgical segmentation, yet it still suffers performance degradation due to downstream domain shift, data corruptions, perturbations, and complex scenes. To further improve its generalization capability and robustness, a broad spectrum of evaluations and extensions remains to be explored and developed. 

\subsubsection{Acknowledgements.}

This work was supported by Hong Kong Research Grants Council (RGC) Collaborative Research Fund (CRF C4063-18G and CRF C4026-21GF), Shun Hing Institute of Advanced Engineering (SHIAE project BME-p1-21) at the Chinese University of Hong Kong (CUHK), General Research Fund (GRF 14203323), Shenzhen-Hong Kong-Macau Technology Research Programme (Type C) STIC Grant SGDX20210823103535014 (202108233000303), and (GRS) \#3110167.

\bibliographystyle{splncs04}
\bibliography{references}

\end{document}